\documentclass[aip,jcp,reprint,amsmath,amssymb,floatfix,citeautoscript]{revtex4-2}
\usepackage{cancel}
\usepackage{amsmath}
\usepackage{mathtools}
\usepackage{physics}
\usepackage{graphicx}
\usepackage{amssymb}
\usepackage{amsthm}
\usepackage{bm}
\usepackage{dcolumn}
\usepackage{braket}
\usepackage{ragged2e}
\usepackage{txfonts}
\usepackage[version=3]{mhchem} 
\usepackage[T1]{fontenc}       
\usepackage[colorlinks=true,allcolors=blue]{hyperref}
\usepackage[capitalise]{cleveref}

\newcommand{\vq}{{\bm{q}}}
\newcommand{\vn}{{\bm{n}}}
\newcommand{\vm}{{\bm{m}}}

\allowdisplaybreaks

\begin{document}

\title{Exact mid-IR quantum vibrational spectra of neutral water clusters}

\author{Henry K. Tran}
\affiliation{Department of Chemistry, Columbia University, New York, New York 10027, USA}
\author{Timothy C. Berkelbach}
\email{t.berkelbach@columbia.edu}
\affiliation{Department of Chemistry, Columbia University, New York, New York 10027, USA}
\affiliation{Initiative for Computational Catalysis, Flatiron Institute, New York, New York 10010, USA}

\begin{abstract}

We use selected configuration interaction to calculate the zero-temperature
mid-infrared (2800--3800~cm$^{-1}$) vibrational spectra of a water monomer, dimer, trimer, and hexamer in
its cage and prism geometries. We use the recently introduced, accurate
q-AQUA-pol potential energy surface along with the $n$-mode representation of
the potential to facilitate grid-based quadrature and integral storage.  Within
selected configuration interaction, we introduce a new approach to the
calculation of spectra that is complementary to eigenstate enumeration.  In the
new approach, we calculate the spectrum using the response-vector method, and
the system of linear equations is solved using a basis of configurations that
are optimally selected at each frequency of interest.
We compare our spectra to previous studies, and highlight limitations of the
local monomer approximation. To the best
of our knowledge, our hexamer spectra are the most accurate ones reported to date.

\end{abstract}

\maketitle

\section{Introduction}
\label{sec:introduction}

Small water clusters are important in atmospheric chemistry, and 
they form a foundation for our understanding of condensed-phase water,
including its atomic structure, hydrogen bond network, and vibrational
dynamics~\cite{Liu1996,Keutsch2001,Ludwig2001}.  In principle, such small
clusters can be studied experimentally via high-resolution gas-phase
spectroscopy following supersonic expansions.  Charged water clusters have been
extensively studied, providing insight into the nature of an excess
proton~\cite{Shin2004,Miyazaki2004,Chang2005} or electron in water.
However, neutral water clusters are significantly more difficult to study due
to the challenge of applying mass spectrometry for size
selection~\cite{Page1984,Huang1989,Pribble1994,Huisken1996,Diken2003,Zhang2020}.
To overcome this limitation, infrared (IR) spectra of neutral water clusters
are typically obtained indirectly as action spectra---for example, in
vibrational predissociation spectroscopies, which measure the depletion of an
ion yield following IR excitation.  Alternatively, neutral water clusters can
be studied in rare gas solids or helium nanodroplets~\cite{Froechtenicht1996},
but the artificial confinement can modify the cluster properties.

Given these experimental challenges, and the fact that supersonic expansions produce
nonthermal distributions, the assignments of the obtained spectra are extremely challenging.
Therefore, calculated IR spectra of low-energy isomers of neutral water clusters
are a powerful, complementary tool for interpretation of these experiments~\cite{Wang2012, Brown2017}.
However, because similarly sized water clusters and their isomers have
very similar spectra, the calculated spectra must be of extremely high accuracy
for correct discrimination and assignment, which limits the applicability of standard
computational methods.

The accuracy of calculated vibrational spectra is limited by two factors: the
employed potential energy surface (PES) and the description of the nuclear
dynamics.  The first limitation is being rapidly overcome through the fitting
or training of PESs to accurate quantum chemistry calculations~\cite{Bukowski2008,
Wang2009, Wang2011a, Babin2012, Babin2013, Yu2022, Qu2023}.
The second limitation is not problematic for molecules containing
heavy atoms at or above room temperature, for which classical nuclear
dynamics is sufficient. For molecules containing light atoms,
the low-temperature dynamics require a quantum mechanical treatment~\cite{Rossi2014}.
Although path-integral methods have proven powerful, they contain
uncontrolled approximations that limit their accuracy at low
temperature~\cite{Jang1999,Habershon2013,Kapil2020}.
Exact quantum spectra can instead be obtained from variational wavefunction
techniques~\cite{Bowman2008}---such as vibrational configuration
interaction~\cite{vscf-1, vscf-2, vscf-ci-1, vscf-ci-2}, tensor network
methods~\cite{vdmrg}, or exact diagonalization with carefully constructed
product bases~\cite{Wang2023,Simko2025}---which are typically limited in the
size of the system they can treat.  In this work, we combine an accurate PES
with new developments in selected vibrational configuration interaction
(VCI)~\cite{hci-1, hci-3, vhci-1, Bhatty2021,Tran2023} to provide numerically exact
quantum spectra of neutral water clusters.

\section{Theory}

In VCI, we calculate vibrational eigenstates
\begin{equation}
|\Psi_\alpha\rangle = \sum_\vn c_\vn^{(\alpha)} |\vn\rangle
\end{equation}
and eigenenergies $E_\alpha$
in a basis of configurations $|\vn\rangle$,
\begin{equation}
\langle \vq | \vn\rangle = \phi_{n_1}(q_1) \phi_{n_2}(q_2) \ldots \phi_{n_N}(q_N),
\end{equation}
where $\phi_{n_i}(q_i)$ are single-mode wavefunctions of mass-weighted
normal modes $q_i$, and the expansion
coefficients $c_\vn^{(\alpha)}$ are eigenvectors of the Hamiltonian matrix in this basis,
$H_{\vm\vn} = \langle \vm | H |\vn\rangle$.  
We use the $J=0$ Watson Hamiltonian,
\begin{equation}
H = -\frac{1}{2} \sum_i \frac{\partial^2}{\partial q_i^2} + V(\vq),
\end{equation}
and we neglect the coupling
between vibrations and rotations, which our testing has shown to be a
good approximation for the large clusters studied here.
Matrix elements of the Hamiltonian require high-dimensional
integration of the PES, which we address using the $n$-mode
representation~\cite{vscf-1, vscf-2},
\begin{equation}
\begin{split}
V(\vq) &= V_0 + \sum_i V^{(i)}(q_i) + \sum_{i<j} V^{(ij)}(q_i,q_j) \\
&\hspace{1em}+ \sum_{i<j<k} V^{(ijk)}(q_i,q_j,q_k) + \ldots, 
\end{split}
\end{equation}
and Gauss-Hermite quadrature~\cite{Harris1965,Dickinson1968}.
We use the one-mode potentials $V^{(i)}(q_i)$ 
to define our single-mode basis functions $\phi_{n_i}(q_i)$,
ensuring that all one-mode anharmonicity is described exactly.

In contrast to traditional VCI, where the basis of configurations $|\vn\rangle$
is truncated by the number of vibrational quanta, we use selected
CI~\cite{Huron1973} to iteratively grow the basis while targeting a manifold of
excited states. The basis is grown by adding configurations $|\vm\rangle$ that
satisfy the so-called heat-bath selection criterion 
$|H_{\vm\vn} c_{\vn}^{(\alpha)}| > \epsilon_1$, where $\epsilon_1$ is a user-selected
threshold with units of energy, and $c_{\vn}^{(\alpha)}$ are the current eigenvectors of
all targeted states $\Psi_\alpha$~\cite{hci-1, hci-3, Holmes2017}.  Importantly, the heat-bath selection
criterion (unlike other selection criteria) does not need to be tested for all
candidate configurations $|\vm\rangle$, as long as the columns of the
Hamiltonian matrix can be implicitly sorted.  In previous
works~\cite{vhci-1,Bhatty2021,Tran2023}, vibrational heat-bath CI (VHCI) has been
used with a Taylor series expansion of the PES, and the current work is the
first to use the more accurate $n$-mode representation.  Briefly, we only
consider one- and two-mode excitations in the selection procedure, by explictly
testing all possible one-mode excitations and by using a pre-sorted list of
integrals $V^{(ij)}_{m_i m_j n_i n_j}$ for two-mode excitations. Hamiltonian
matrix elements are always calculated using the full $n$-mode representation of
the PES.  Further details of this extension are given in the Supporting
Information. In principle, the HCI results can be improved by second-order
perturbation theory~\cite{hci-1,hci-3,Holmes2017}, but this does not work well in highly excited regions
of the spectrum~\cite{vhci-1}, so we do not pursue it here.

At each iteration of VHCI, the basis grows and therefore we obtain more accurate
estimates of the eigenstates and energies.
The eigenstates and eigenenergies can then be used to calculate the $T=0$ infrared
spectrum in a sum-over-states (SOS) fashion,
\begin{equation}
I(\omega) = \sum_{a=x,y,z} \sum_{\alpha > 0} |\langle \Psi_\alpha | \mu_a | \Psi_0\rangle|^2
    \delta[\omega-(E_\alpha-E_0)]
    \equiv \sum_{a=x,y,z} I_{aa}(\omega).
\end{equation}
where the dipole moment surface $\mu_a(\vq)$ has an $n$-mode representation to
facilitate integration.
However, for large clusters with high-dimensional PESs, the enumeration of all
eigenstates contributing to the spectrum can be prohibitive due to the high density of
states. 
Within HCI, targeting many eigenstates will cause the basis to grow quickly,
which can make it hard to obtain a converged spectrum.
Therefore,
we introduce a new selected CI method that is complementary to eigenstate
enumeration. The new method directly calculates the spectral
intensity at arbitrary frequencies.
Using the Lorentzian representation of the delta function with linewidth $\eta$,
we write the spectrum as
\begin{equation}
\begin{split}
I_{aa}(\omega) 
    &= -\pi^{-1} \mathrm{Im} \langle \Psi_0 | \mu_a \left[\omega-(H-E_0)+i\eta)\right]^{-1}\mu_a|\Psi_0\rangle \\
&= -\pi^{-1} \mathrm{Im} \langle \Psi_0 | \mu_a | X_a(\omega) \rangle
\end{split}
\end{equation}
where $|X_a(\omega)\rangle$ is the response vector (also known as the correction vector)
that solves the linear equation~\cite{Svendsen1985,Soos1989,Kuehner1999}
\begin{equation}
\label{eq:Axb}
[\omega-(H-E_0)+i\eta]|X_a(\omega)\rangle = \mu_a|\Psi_0\rangle.
\end{equation}
Following the selected CI method for eigenstates,
we expand the unknown solution vector in a basis of configurations,
\begin{equation} \label{eq:response_vector}
|X_a(\omega)\rangle = \sum_{\vn} x_{\vn}(\omega) |\vn\rangle,
\end{equation}
and we generate the basis iteratively.  First, on the basis of
Eq.~(\ref{eq:Axb}), we add configurations $|\vm\rangle$ satisfying the
heat-bath style criterion $|\left(\mu_a\right)_{\vm\vn} c_\vn^{(0)}| > \mu_1$, where
$\mu_1$ is a threshold with units of the dipole moment.  Second, we add
configurations that are tailored to the frequency of interest $\omega$: as
shown in the SM, perturbation theory suggests that the basis should be grown by
adding configurations that satisfy $|H_{\vm\vn} \mathrm{Im} x_\vn(\omega) | >
\mu_2$, where $\mu_2$ is a second threshold.  Importantly, this latter step
only adds configurations that are important for resolving the spectral
intensity near the target frequency $\omega$, and spectral intensities at each
$\omega$ can be calculated independently and in parallel.  Aside from the
ground state, no eigenstates are explicitly calculated.
In this manuscript, we will compare the SOS VHCI approach and the spectral VHCI
approach, both of which can in principle be converged to exactness.

\section{Computational Details}
We now apply these methods to neutral clusters of water molecules, namely the
dimer, trimer, and hexamer.
We use the q-AQUA-pol
PES~\cite{Yu2022,Qu2023}, which is the polarizable TTM-3F
potential~\cite{Fanourgakis2008} corrected by two-, three-, and four-body terms
fitted to an extensive dataset of CCSD(T) calculations.  The harmonic
frequencies of water hexamers calculated with the q-AQUA-pol potential agree
with those of benchmark CCSD(T) calculations~\cite{Howard2015} with a mean
absolute deviation of only 5~cm$^{-1}$, suggesting that the potential is very
reliable for vibrational frequency calculations of the neutral water clusters
studied here.  
We use the dipole moment surface (DMS) of Ref.~\onlinecite{Liu2015}, which
includes the spectroscopically accurate one-body LTP2011 DMS~\cite{Lodi2011}
supplemented by two-body terms fitted to a dataset of MP2 calculations.  This
DMS was shown to yield accurate IR intensities for a water dimer~\cite{Liu2015}
and for liquid water~\cite{Liu2015,Liu2015a}.

To facilitate integration, we use a three-mode representation of the q-AQUA-pol
PES. Consistent with previous work~\cite{Wang2008,Yang2023}, we find that this
truncation of the potential, and its approximation to four-mode couplings and
higher, leads to spurious minima or unbound potentials. Low-frequency torsional
modes are especially problematic when using rectilinear
coordinates~\cite{Wang2008}.  Therefore, in our calculations, we freeze all
low-frequency modes, focusing on the vibrational signatures of the water
stretches and bends (three modes per molecule); this is the same approximation
made in pioneering VCI studies of the water dimer and trimer~\cite{Wang2008}.
Our water dimer, trimer, and hexamer calculations are thus 6-, 9-, and
18-dimensional VCI calculations.  For the water dimer and trimer, we have
confirmed that the high-frequency spectra are converged with the three-mode
representation of the PES.  In all spectra, we use a Lorentzian linewidth (half
width at half maximum) of $\eta = 25$~cm$^{-1}$.  Full computational details
are provided in the Supporting Information.

\begin{figure*}[t]
	\includegraphics[scale=1.0]{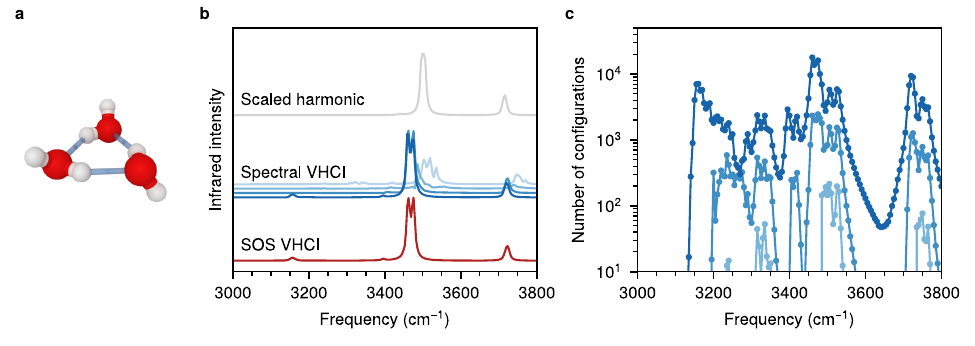}
    \caption{(a) Geometry of the minimum energy up-up-down water trimer. (b)
Simulated IR spectrum of the water trimer in the scaled harmonic approximation
(grey), spectral VHCI for various thresholds $\mu_2$ (blue), and by summing
over states (SOS) calculated by VHCI (red).  The VHCI ground state was
performed with $\epsilon_1 = 10$~cm$^{-1}$, and the spectra were calculated with
$\mu_1 = 2.5$~D and $\mu_2 =  2.5\times 10^{-n}$~D with $n=0,1,2,3$. 
    (b) Number of configurations added to the ground state variational space to
perform spectral VHCI at each frequency, with the same thresholds shown in (a).
}
	\label{fig:trimer}
\end{figure*}

\section{Results}

\begin{table}[b]
    \centering
    \caption{ZPE and fundamental excitation energies for the water dimer (energies in cm$^{-1}$)
    using the q-AQUA-pol PES and the HBB PES (results from Ref.~\onlinecite{Wang2008}).
    The six lowest frequency modes, with a harmonic ZPE of 779.0~cm$^{-1}$, were frozen in all calculations.}
    \label{tab:dimer}
    \begin{tabular*}{\columnwidth}{@{\extracolsep{\fill}} lccccccc}
        \hline\hline
        & ZPE & $\omega_1$ & $\omega_2$ & $\omega_3$ & $\omega_4$ & $\omega_5$ & $\omega_6$ \\
Harmonic& 9373.8 & 1650.0 & 1670.2 & 3748.3 & 3828.1 & 3916.1 & 3934.7 \\
VCI     & 9240.2 & 1582.1 & 1601.7 & 3548.3 & 3652.1 & 3718.9 & 3735.0 \\ 
VCI (HBB)&       & 1589.4 & 1604.2 & 3548.8 & 3636.7 & 3700.7 & 3722.2 \\ 
        \hline\hline
     \end{tabular*}
 \end{table}          

\begin{table*}[t]
    \centering
    \caption{ZPE and fundamental excitation energies for the water trimer (energies in cm$^{-1}$)
    using the q-AQUA-pol PES and the HBB PES (results from Ref.~\onlinecite{Wang2008}).
    The nine lowest frequency modes, with a harmonic ZPE of 2181.5~cm$^{-1}$, were frozen in all calculations.}
    \label{tab:trimer}
    \begin{tabular*}{\linewidth}{@{\extracolsep{\fill}} lcccccccccc}
        \hline\hline
        & ZPE & $\omega_1$ & $\omega_2$ & $\omega_3$ & $\omega_4$ & $\omega_5$ & $\omega_6$ & $\omega_7$ & $\omega_8$ & $\omega_9$\\
Harmonic& 13867.0 & 1662.1 & 1665.3 & 1684.0 & 3621.0 & 3681.1 & 3689.2 & 3907.0 & 3910.1 & 3914.2 \\
VCI     & 13649.4 & 1591.3 & 1595.2 & 1615.0 & 3394.3 & 3460.4 & 3473.8 & 3716.3 & 3720.2 & 3723.7 \\
VCI (HBB)&        & 1646   & 1659   & 1674   & 3463   & 3533   & 3544   & 3750   & 3754   & 3765   \\ 
        \hline\hline
     \end{tabular*}
 \end{table*}

The water monomer and dimer are small and their spectra are easily calculated
exactly.  For the monomer, the q-AQUA-pol potential reduces to that of
Partridge and Schwenke~\cite{Partridge1997}, and the three-mode representation
of the potential is exact.  For the monomer, we calculate a ZPE of
4648.0~cm$^{-1}$ and three fundamental excitation energies of 1581.7 cm$^{-1}$,
3656.1 cm$^{-1}$, and 3741.9 cm$^{-1}$ (these differ by up to 15~cm$^{-1}$ from
exact $J=0$ values~\cite{Partridge1997,Huang2005} due to our neglect of
rotation-vibration coupling).  For the dimer, the ZPE and fundamental
excitation energies are given in Tab.~\ref{tab:dimer}.  Compared to previous
VCI results obtained with the HBB potential~\cite{Huang2008,Wang2008} but
otherwise similar approximations, our excitation energies differ by 10--20
cm$^{-1}$, demonstrating the impact of recent PES developments.

We now turn to the trimer, a 9-dimensional problem. 
We study the lowest-energy `up-up-down' isomer, the geometry of which is
shown in Fig.~\ref{fig:trimer}a.
From a converged VHCI ground state, we calculate a ZPE of
15831~cm$^{-1}$, for which we used the harmonic ZPE of the 12
low-frequency modes (2182~cm$^{-1}$); this number is in reasonable agreement with the value of
15616$\pm$2~cm$^{-1}$ obtained for the full q-AQUA-pol potential using
diffusion Monte Carlo~\cite{Qu2023}, although it is hard to assign the
discrepancy to the three-mode representation of the potential
or the harmonic treatment of low-frequency modes. 
We expect excitation energies to be significantly more accurate
due to cancellation of errors.

With low-frequency modes frozen, the trimer is small enough to allow an exact
calculation by brute-force SOS VHCI,
which allows us to benchmark our new spectral VHCI method.
The calculated IR spectrum is 
plotted in Fig.~\ref{fig:trimer}(a), comparing the scaled 
harmonic approximation (scaling factor 0.95),
the exact SOS VHCI result, 
and the spectral VHCI result using various thresholds $\mu_2$.
We observe that with increasingly tight thresholds, spectral VHCI 
converges to the exact spectrum as expected.
The converged spectra show two intense peaks at 3460 and 3474~cm$^{-1}$,
which are predominantly hydrogen-bonded OH stretching modes.
After scaling, the harmonic approximation captures these peaks quite well (they
appear as a single peak in Fig.~\ref{fig:trimer}(a) due to broadening).
The VHCI spectra show a small feature around 3150~cm$^{-1}$, which is a bending
overtone that is of course absent in the harmonic spectrum.

The advantage of spectral VHCI is its lower computational cost, which is
attributable to the tailored selection of configurations at each frequency. In
Fig.~\ref{fig:trimer}(b), we show the final number of configurations used to
calculate the spectrum at each frequency, averaged over the $x$, $y$, and $z$
components.  Importantly, a large number of configurations is only required
when there is a peak in the spectrum. In this particular calculation, the exact
spectrum was calculated by summing over 20 eigenstates calculated
in a selected CI basis of about 5100 configurations, which was found sufficient
to converge the spectrum to graphical accuracy. The spectral
VHCI calculation achieves the same graphical convergence using the second-tightest
threshold shown ($\mu_2 = 2.5\times 10^{-2}$~D), which required, at most, about 2500 configurations
near 3500~cm$^{-1}$, and less than 1000 configurations at almost all other frequencies.
Both the
eigenvalue calculation and the linear solve build the Hamiltonian matrix in
this basis, such that the smaller number of configurations yields a storage
reduction of a factor of four or more.  Moreover, the time required to solve
the linear equation is signficantly reduced in a smaller basis, whether
using an exact linear solver, such as LU decomposition, or an iterative
linear solver, such as the generalized minimum residual (GMRES) method. 
Although exact timings depend on implementation details, our pilot
implementation of the spectral VHCI algorithm required about ten minutes for
the most expensive frequency points, and significantly less for most frequency
points.

\begin{figure*}[t]
	\includegraphics[scale=1.0]{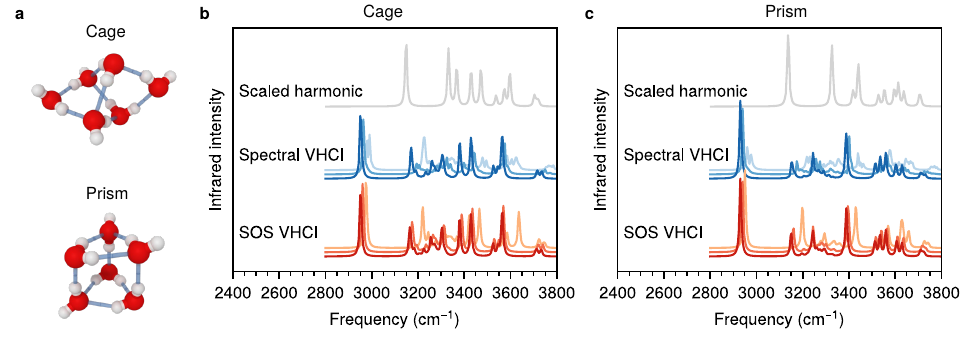}
    \caption{(a) Geometry of the cage and prism water hexamers. (b) and (c)
Simulated IR spectrum of the cage (b) and prism (c) hexamer in the scaled
harmonic approximation (grey), spectral VHCI for various thresholds $\mu_2$
(blue), and by summing over states (SOS) calculated by VHCI with
$\epsilon_1=10^{-1}$~cm$^{-1}$ after the first, second, and third iterations
(red). In the spectral VHCI calculation, the ground state was calculated with
$\epsilon_1 = 10$~cm$^{-1}$, and the spectra were calculated with ($\mu_1$,
$\mu_2$) = ($2.5\times 10^{-m}$~D, $2.5\times 10^{-n}$~D) with $(m,n) = (4,0),
(5,1), (6,2)$. 
}
	\label{fig:hexamer}
\end{figure*}

As a more challenging demonstration of our methods, we simulate the spectrum of the water hexamer,
which has been extensively studied because it is the smallest cluster with
three-dimensional structure in many of its low-energy isomers.  Although
existing experimental IR spectra, acquired in helium
nanodroplets~\cite{Nauta2000,Burnham2002} and with
argon-tagging~\cite{Diken2003}, have been attributed to the ring and book
isomers, here we study the cage and prism isomers, which are shown in
Fig.~\ref{fig:hexamer}a.  Full-dimensional quantum simulations with an accurate
PES have demonstrated that these two are the lowest energy isomers (with almost
identical energies) and are thus the relevant isomers in equilibrium at low
temperature~\cite{Wang2012,Yu2022}.  Our ground-state VHCI calculations predict
zero-point energies of 32629 and 32678~cm$^{-1}$, which compare well to those
calculated by diffusion Monte Carlo, 32553$\pm$19~cm$^{-1}$ and
32647$\pm$9~cm$^{-1}$ (using the closely related q-AQUA
potential~\cite{Yu2022}), for the cage and prism geometries respectively.

Our calculated IR spectra are shown in Figs.~\ref{fig:hexamer}b and c.  These
18-dimensional quantum calculations are too expensive for an exact calculation,
but we present spectra from increasingly accurate SOS VHCI calculations
(plotting the spectrum after each iteration for fixed $\epsilon_1$) and
spectral VHCI calculations (plotting the spectrum for decreasing $\mu_2$).  The
most accurate spectra from each method show very good agreement for both
isomers, suggesting that they are essentially converged.  The sizes of the
variational spaces suggest that spectral VHCI is especially efficient: the most
accurate spectra required over 65,000 configurations for SOS VHCI, but only about
25,000 configurations for spectral VHCI.
Simultaneously tightening both $\mu_1$
and $\mu_2$ was found to be beneficial in the spectral VHCI results.
We attribute the better performance of spectral VHCI to its
use of dipole intensity information when growing the basis of configurations.
In contrast, the SOS VHCI approach is improving the description of many
eigenstates independent of their intensity, and many of them are
spectroscopically dark or weak.

\begin{figure}[b]
	\includegraphics[scale=1.0]{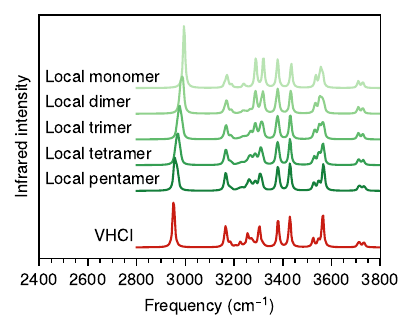}
    \caption{IR spectrum of the water hexamer in its cage geometry, comparing our best VHCI result
to those of the local monomer, dimer, etc., approximations. 
}
	\label{fig:hexamer_lmon}
\end{figure}

As seen in Fig.~\ref{fig:hexamer}, the scaled harmonic spectrum is reasonably accurate at
high frequencies, but inaccurate below about 3500~cm$^{-1}$, due to significant
anharmonic mixing of configurations.
To the best of our knowledge, the previously most accurate spectra of these hexamers 
have all been obtained using the local monomer 
approximation~\cite{Wang2011,Wang2013,Brown2017}.
In this approximation, the vibrational spectrum of each water
monomer is calculated while holding all others fixed at their equilibrium
geometry, and then all monomer spectra are summed~\cite{Wang2011}.  
In Fig.~\ref{fig:hexamer_lmon}, we compare our own local monomer spectra for
the q-AQUA-pol PES with our best VHCI spectrum.  Our local monomer spectra are
graphically similar to ones previously calculated with the MB-pol
PES~\cite{Brown2017}, although ours are shifted to lower frequencies by about
40~cm$^{-1}$.  For simplicity, we only show results for the cage geometry;
results for the prism geometry are qualitatively similar and given in the SI.

In contrast to the inaccurate scaled harmonic spectra, the local monomer
spectra are in reasonable agreement with those from spectral VHCI,
but they miss some of the fine structure, especially around 3200--3300~cm$^{-1}$.
This fine structure is due to many multiconfigurational states, due to anharmonic mixing between
OH stretches and overtones of highly delocalized bending modes.
The largest discrepancy is in the
position of the intense, low-frequency peak around 2950--3000~cm$^{-1}$. 
The VHCI peak is about 50~cm$^{-1}$ lower than the
local monomer peak. 
This transition is essentially a single
OH stretch along one of the hydrogen bonds whose accepting molecule has no free
hydrogens.
This discrepancy, even for a seemingly simple 
intramolecular excitation, demonstrates the challenge of achieving
quantitative accuracy in the spectroscopy of high-dimensional systems.

To explore the errors of the local monomer approximation,
we performed analogous local dimer, trimer, tetramer, and pentamer
calculations, by averaging the spectra from fifteen 6-dimensional calculations,
twenty 9-dimensional calculations, fifteen 12-dimensional calculations, and six
15-dimensional calculations.  The reference VHCI calculations, which are
18-dimensional, go beyond all of these calculations by including up to
three-mode coupling between all water monomers.  The fine structure around
3200--3300~cm$^{-1}$ is resolved reasonably correctly with the local trimer
approximation, which we attribute to the delocalized character of the bending overtones
that mix in this region. The lowest-frequency peak continuously shifts to lower
frequency and is not correct until the full hexamer calculation.

\section{Conclusion}
\label{sec:conclusion}

We have combined recent developments in PES parameterization and selected
vibrational configuration interaction to study the IR spectrum of neutral water
clusters. To the best of our knowledge, our mid-IR spectra of the hexamers are
the most accurate ones reported to date. We hope our results
serve as a valuable benchmark to the cost-effective local monomer approximation,
as we have shown, and perhaps to classical or semiclassical methods, like those
based on path integrals.

Although VHCI alleviates the exponential scaling of quantum simulations, the
method is now bottlenecked by the costs of quadrature and integral handling.
These costs are responsible for the most severe approximation of the present
work, i.e., the use of the three-mode representation of the PES, which also
requires that we freeze low-frequency modes to avoid unphysical behavior.
Alternative PES representations can lower these costs while simultaneously suggesting
new, faster configuration selection schemes, as discussed in our previous work~\cite{Tran2023}.
With these improvements, the methods described here can be straightforwardly applied
to problems with more degrees of freedom, and work along these lines is in progress.

\section*{Acknowledgements}
We thank Qi Yu for sharing code for the dipole moment surface of water used in
this work.  T.C.B.~thanks Joel Bowman for many helpful discussions, as well as
Ankit Mahajan and Sandeep Sharma for early discussions about spectral functions
with HCI.  This work was supported by the U.S. Department of Energy, Office of
Science, Basic Energy Sciences, under Award No.~DE-SC0023002.  We acknowledge
computing resources from Columbia University's Shared Research Computing
Facility project, which is supported by NIH Research Facility Improvement Grant
1G20RR030893-01, and associated funds from the New York State Empire State
Development, Division of Science Technology and Innovation (NYSTAR) Contract
C090171, both awarded April 15, 2010.  The Flatiron Institute is a division of
the Simons Foundation.

\end{document}